\documentclass[english,aps,prb,a4paper,floatfix]{revtex4}
\usepackage[T1]{fontenc}
\usepackage{tikz}
\usepackage{color}
\usepackage{amsmath}
\usepackage{amssymb}
\usepackage[latin1]{inputenc}
\usepackage{graphicx}
\usepackage{bm}
\usepackage{epsfig}
\usepackage{xcolor}
\usepackage{tikz,pgfplots}
\usepackage{tcolorbox}
\usepackage{float}
\usepackage{verbatim}
\usetikzlibrary{positioning}
\definecolor{mynicegreen}{RGB}{102,182,102}

\newcommand{\hypgeo}[2]{%
  \operatorname{%
    {\vphantom{\mathnormal{F}}}_{#1}%
    \kern-\scriptspace
    \mathnormal{F}_{#2}%
  }%
}
\providecommand{\keywords}[1]
{
  \small	
  \textbf{\textit{Keywords---}} #1
}

\begin{document}

\title{Prediction of the liquid-crystal phase behavior of hard right triangles from fourth-virial density-functional theories}
\author{Enrique Velasco}
\email{enrique.velasco@uam.es}
\affiliation{Departamento de F\'{\i}sica Te\'orica de la Materia Condensada,
Instituto de F\'{\i}sica de la Materia Condensada (IFIMAC) and Instituto de Ciencia de
Materiales Nicol\'as Cabrera,
Universidad Aut\'onoma de Madrid,
E-28049, Madrid, Spain}

\author{Yuri Mart\'{\i}nez-Rat\'on}
\email{yuri@math.uc3m.es}
\affiliation{
Grupo Interdisciplinar de Sistemas Complejos (GISC), Departamento
de Matem\'aticas, Escuela Polit\'ecnica Superior, Universidad Carlos III de Madrid,
Avenida de la Universidad 30, E-28911, Legan\'es, Madrid, Spain}

\date{\today}

\begin{abstract}
We have used an extended Scaled-Particle Theory that incorporates four-body
correlations through the fourth-order virial coefficient to analyse the
orientational properties of a fluid of hard right-angle triangles. This fluid
has been analysed by computer simulation studies, with clear indications of strong
octatic correlations present in the liquid-crystal phase, although the more symmetric 
order tetratic phase would seem to be the most plausible candidate. Standard theories
based on the second virial coefficient are unable to reproduce this behaviour.
Our extended theory predicts that octatic correlations, associated to a symmetry
under global rotations of the oriented fluid by $45^{\circ}$, are highly
enhanced, but not enough to give rise to a thermodynamically stable phase
with strict octatic symmetry. We discuss different scenarios to
improve the theoretical understanding of the elusive octatic phase in this
intriguing fluid.

\end{abstract}


\maketitle

\section{Introduction}
\label{intro}
Fluids of elongated particles in two dimensions (2D) \cite{deGennes} continue to unveil surprising
behaviours \cite{Musevic}. Since the discovery of crystallisation in the hard-disc systems,
fluids made of hard elongated particles, such as ellipsoids, have been seen
to stabilise nematic phases with uniaxial symmetry \cite{Frenkel}. As common in 2D,
these phases possess quasi-long-range orientational order \cite{Frenkel1,Cuesta,Odriozola},
but mean-field theories can still describe large regions where the
tensor order parameter exhibits persistent values \cite{Kleman,Wensink}. Since particles interact through purely
overlap interactions, order in 2D phases made of hard elongated particles 
is solely governed by entropy, which shows here its most subtle nature.

More complicated 2D particle shapes have been explored more recently from a
theoretical perspective \cite{Schlacken,MAR0,MAR,Schmidt,Triplett,Wittmann,Escobedo0,Sabi,Quintana,Lowen,cinacchi}, motivated by the possibility to fabricate colloidal
particles of virtually any shape\cite{exp1,exp2,exp3}. Particles with regular polygonal shapes
have been demostrated to exhibit mesophases \cite{Schlacken,Anderson}, with $4$-atic (or tetratic,
with two directors and a global symmetry under rotation by $90^{\circ}$)
phases for squares, $6$-atic (with three directors and a symmetry
under $60^{\circ}$) for equilateral triangles and hexagons.
No further mesophases appear to get stabilised for polygonal shapes
with more edges: they crystallise directly from the isotropic fluid through
a KTHN-type transition, as in the case of discs \cite{Anderson,Mak}. Simulation of mixtures of
particles have also found interesting behaviours \cite{Escobedo}. 

Nonregular polygons open up new possibilities \cite{MAR,dani}. Here entropy plays an even more subtle role:
particles tend to form local clusters of oriented particles that can
be viewed as `superparticles', with symmetries sometimes very different
from that of the `monomers' and therefore from the symmetry of the
bulk liquid-crystal phase that would trivially follow from the monomers. 
Such is the case in fluids made of low-aspect-ratio rectangles, 
which tend to form highly stable square clusters that stabilise
a global $4$-atic phase \cite{Granada}. The same behaviour is found in vibrated
monolayers of granular grains \cite{Muller,granos_nosotros} and in experiments on colloidal particles
\cite{exp3}. Hard-kite shaped particles have been studied by simulation \cite{sim} and theory
\cite{kites}, and $4$-atic phases were also found. The basic understanding of this phase lies in the
excluded area between particles (second-order virial coefficient), an essential
ingredient of the Scaled-Particle Theory (SPT) extension of classical 
Onsager theory. Three-body correlations can be incorporated into the theory through the third-order virial
coefficients \cite{Mederos}, and the ensuing corrections are important:
the stability region of the $4$-atic phase is extended to larger aspect
ratios and lower densities.

Recently we have studied a fluid made of hard right-angle isosceles triangles (HRT) \cite{MAR1}. 
Motivated by Monte Carlo (MC) simulations by Gantapara et al. \cite{Gan}, we analysed
the fluid using the standard SPT, based on the second virial
coefficient (which is analytic), and an extension that includes the third-virial 
coefficient, calculated using MC integration.
It turns out that none of these theories can reproduce the behaviour predicted
from the simulations: as the isotropic fluid is compressed,
clustering of particles in clusters of various shapes give rise to
strong $8$-atic correlations, and an orientational distribution function
with $4$-atic symmetry but high secondary peaks at $45^{\circ}$ with
respect to the main peaks results. The equilibrium orientational function
from the theories, by contrast, shows no hint of the high-order
$8$-atic symmetry.

The HRT fluid seems to be the first case where an Onsager-type
theory fails to give even a qualitative picture of a mesophase. In previous
work we have discussed this problem. On the one hand, the theory is strictly
valid for infinitely long rods, while here we are considering particle of low
aspect ratio. On the other, the sequence of virial coefficients in 2D fluids is known to 
have peculiar behaviour, and the condition that scaled higher-order coefficients
are small is not fulfilled. Finally, a crucial property of the HRT fluid and other
2D fluids made of hard polygonal particles is the strong clustering tendency
of the fluid \cite{MAR,MAR1}, with at least five types of clusters that
form in the fluid at intermediate phases (before crystallisation). Four
of these clusters involve two particles (dimers), of square, triangular
and rhomboidal shapes (with two enantiomers in the latter case), and
one involve four particles arranged in a square. The fluid can somehow be
viewed as a multicomponent mixture of dynamic `superparticles' and might
be more quantitatively described by association theories than by particle
(monomer)-based theories. Before undertaking such a programme, we speculated
\cite{MAR,MAR1} that a theory based on four-body correlations (i.e. on the fourth virial
coefficient) might give some indication as to whether high-order particle
correlations, involved in clustering tendencies of the particles, might
be important to understand the equilibrium structure of the fluid.

In the present work we show the predictions of such a theory. A resummed SPT
is developed using the standard technique \cite{Padilla,Mederos,MAR1}, 
which allows to systematically incorporate an arbitrary
number of virial coefficients. These objects are generalised virial coefficients
in the sense that they are functionals of the orientational distribution function.
The third and fourth virial coefficients are computed numerically, and the 
instability of the isotropic (I) phase against orientational orders of different symmetries
is investigated. This process allows to analyse the effect of increasing low-order,
from two- to four-particle correlations on the onset of bulk orientational order. 
Focusing on the $8$-atic (or octatic)
orientational symmetry, we explore the tendency of the fluid to stabilise orientational
order through a bifurcation analysis. Our conclusion is that four-particle correlations
do enhance octatic symmetry. More definite conclusions as to the orientational distribution
of particles would require a full minimization of the free energy. However, a more quantitative
theory should incorporate particle clustering \cite{MAR1}, which is not possible with the present SPT
scheme, and further studies will have to await until a proper treatment of clusters can
be formulated.

The article is organised as follows. In Section \ref{theory} we present
the theory and a method to extend SPT to include the fourth virial
coefficient. Also, we provide some details on the numerical calculation
of this coefficient. In Section \ref{results} we present the results,
along with the results obtained from extrapolations of virial
coefficients and resummations of the virial theory in the isotropic
phase, which may help to explain the role of many-particle correlations.
Some conclusions are drawn in Section \ref{conclusions}. The Appendix
collects some further numerical details and additional results.

\section{Theory}
\label{theory}

Our theory uses the same strategy followed in our previous works on the
third virial coefficient \cite{Mederos,MAR1}, but now the theory is
extended to an arbitrary number of virial coefficients.
We start from the following analytic expression for the equation of state, valid
for a fluid of oriented particles:
\begin{eqnarray}
	&&\beta p a=\frac{\eta}{1-\eta}+
	\frac{\displaystyle\sum_{k=2}^n c_k[h]\eta^k}{\displaystyle(1-\eta)^q}.
	\label{lolo} 
\end{eqnarray}
The parameter $q$ in Eqn. (\ref{lolo}) controls the divergence of the second term at close-packing, $\eta=\eta_{\rm cp}\equiv 1$. It is taken 
as a free parameter and will be important 
to compare the resulting equation of state with MC simulations. 
$\eta=\rho a$ is the packing fraction, with $\rho$ and $a$ 
the number density and particle area, respectively. $h(\phi)$ is the 
orientational distribution function, which quantifies the average orientation of the 
particles, where $\phi$ is the angle between the averaged orientation of the main
symmetry axis of a typical particle and the main director. This function is normalised,
\begin{eqnarray}
        \int_0^{2\pi}d\phi h(\phi)=1.
\end{eqnarray}
The coefficients $c_n[h]$ are functionals of the orientational distribution function
and can be related to the standard virial coefficients 
by equating the low-density expansion of (\ref{lolo}) with the exact virial expansion 
up to $n$th order (note that the expansion in Eqn. (\ref{lolo}) is 
truncated at this order):
\begin{eqnarray}
	c_n[h]=b_n[h]+\sum_{k=1}^{n-2}(-1)^k\binom{q}{k}b_{n-k}[h], 
\end{eqnarray}
where we have defined 
\begin{eqnarray}
	b_n[h]\equiv \frac{B_n[h]}{a^{n-1}}-1,
\end{eqnarray}
with $B_n[h]$ the standard $n$th virial coefficient. Note that these coefficients
are also functionals of $h(\phi)$. 
Fixing $q=0$ in Eqn. (\ref{lolo}) the exact virial expansion up to the $n$th order is recovered. 
Also, selecting $q=2$ and $n=2$ we obtain the SPT approximation, while 
for the same $q$ and $n=3$ the equation of state turns out to be the same as the
one used in \cite{MAR1} to study fluid orientational ordering close to the 
I-$p$--atic bifurcation point, henceforth called $B_3$-SPT. This shows that
Eqn. (\ref{lolo}) is a versatile starting point to explore nonstandard 
approximations to the equation of state of oriented fluids in two dimensions.

In this article we start by exploring the case $q=2$ and $n=4$ (henceforth
called $B_4$-SPT), and then use extrapolated values for the first three
virial coefficients, $B_2$, $B_3$ and $B_4$ of the I phase to predict the value of $B_5$,
giving rise to the $B_5^*$-SPT theory (the asterisk denotes an extrapolated value for $B_5$). 
Not only the values of the virial coefficients
for the completely disordered fluid are calculated, but also relevant Fourier components
with respect to weak orientational order of particular $p$-atic symmetries. This allows for a 
bifurcation analysis of the I phase with respect to these symmetries.
Finally, the effect of $q$ has also been assessed for $n=3$, $4$ and $5$, which can be
used to explore the effect of the divergence of the equation of state on the 
relative locations of the I-$2$--atic and I-$8$-atic bifurcations, and in general on
the performance of the theory when compared with MC simulations. 

In order to proceed we first need to derive the free energy. As usual it is more convenient
to obtain the excess free energy by integrating the excess pressure, which can be written
as $\beta p_{\rm exc}a =\beta p a-\eta$ ($\beta p_{\rm id}=\rho$ is the ideal pressure). 
Using the thermodynamic relation 
$\displaystyle{\beta p_{\rm exc} a =\eta^2\frac{\partial \varphi_{\rm exc}}{\partial \eta}}$, with 
$\varphi_{\rm exc}$ the excess part of the Helmholtz free-energy per particle, we can integrate 
the expression above with respect to $\eta$ to obtain
\begin{eqnarray}
	\varphi_{\rm exc}[h]=-\log(1-\eta)+\sum_{k=2}^n c_k[h]\hypgeo{2}{1}(k-1,q;k;\eta)\frac{\eta^{k-1}}{k-1}.
\end{eqnarray}
The integration constant was set to zero to ensure that the excess free energy be zero at the 
low-density limit. $\hypgeo{2}{1}$ is the hypergeometric function,
\begin{eqnarray}
	\hypgeo{2}{1}\left(n,q;n+1;\eta\right)\frac{\eta^n}{n}=\int_0^{\eta} \frac{u^{n-1}}{(1-u)^q}du.
\end{eqnarray}
The ideal part of the free-energy per particle has the exact form
\begin{eqnarray}
	\varphi_{\rm id}[h]=\log \eta -1 +\int_0^{2\pi} d\phi h(\phi) \log\left(2\pi  h(\phi)\right),
\end{eqnarray}
giving the total free-energy functional as $\varphi[h]=\varphi_{\rm id}[h] + \varphi_{\rm exc}[h]$.

Since we will only be interested in assessing the contribution of higher-order correlations
to the stability of the I phase against $p$-atic symmetries, we restrict here to a stability 
analysis of the functional against orientational fluctuations of a given symmetry. 
In practice this means that the full Fourier expansion of $h(\phi)$ can be truncated,
keeping only the term with the required symmetry. More specifically, close
to the I-$2m$-atic ($m=1,\dots,4$) bifurcation point, we approximate 
\begin{eqnarray}
	h(\phi)\simeq \frac{1}{2\pi}\left[1+h_m \cos(2m\phi)\right],
\end{eqnarray}
with $h_m$ the first Fourier amplitude with $m$-th symmetry. Explicitely, $m=1$ (2$-$atic), 
$2$ (4$-$atic), $3$ (6$-$atic) and $4$ (8$-$atic).
Inserting this expression in the total free energy functional $\varphi[h]$ we obtain, to lowest
order in $h_m$,
\begin{eqnarray}
\varphi[h]=\varphi_{\rm I}+\Delta \varphi[h], 
\end{eqnarray}
where
\begin{eqnarray}
\varphi_{\rm I}=\log \left(\frac{\eta}{1-\eta}\right)-1
	+\sum_{k=2}^n {\cal C}_k^{(0)}[h]\frac{\eta^{k-1}}{k-1}\hypgeo{2}{1}(k-1,q;k;\eta),
\end{eqnarray}
is the free energy of the I phase, and
\begin{eqnarray}
\Delta\varphi[h]=\frac{h_m^2}{2}\chi_m(\eta), \ m\geq 1
\end{eqnarray}
with
\begin{eqnarray}
\chi_m(\eta)=1+2\sum_{k=2}^n \frac{{\cal C}_k^{(m)}\eta^{k-1}}{(k-1)!} \hypgeo{2}{1}(k-1,q;k;\eta),\quad m=1,\dots,4,
\end{eqnarray}
is the extra contribution associated to an orientational fluctuation with symmetry of order $m$ and
amplitude $h_m$. In the expressions above we have defined
\begin{eqnarray}
	{\cal C}_n^{(m)}={\cal B}_n^{(m)}+\sum_{k=1}^{n-2}(-1)^k\binom{q}{k}{\cal B}_{n-k}^{(m)},
\end{eqnarray}
while the coefficients ${\cal B}_n^{(m)}$ denote the Fourier components of the scaled virial coefficients.
To quadratic order
\begin{eqnarray}
	b_n[h]={\cal B}_n^{(0)}+\frac{h_m^2}{2}{\cal B}_n^{(m)}.
\end{eqnarray}
The procedures used to calculate the ${\cal B}_n^{(m)}$ coefficients are described in Appendix \ref{app}, while
the numerical values for the bifurcations are discussed in Sec. \ref{results}. We only note here that, in order to
find the value of packing fraction $\eta_m$ for the I-$2m$-atic bifurcation we need to solve the equation 
$\chi_m(\eta)=0$. 

In this work we limit the maximum order of the virial coefficient to $m=4$. However, it will be worthwhile to
explore the effect of an extrapolated fifth and possibly higher-order virial coefficients. From Eqn. (\ref{lolo}) 
with $n=4$, i.e. the $B_4$-SPT approximation, we can obtain the relation between 
$b_m[h]$ for $m\geq 5$ and $b_2[h]$, $b_3[h]$ and $b_4[h]$:
\begin{eqnarray}
	b_m[h]&=&\left\{2(m-2)(m-3)b_4[h]-2(m-2)(m-4)(q-1)b_3[h]
	\right.\nonumber\\
	&+&\left.(m-3)(m-4)(q-1)(q-2)b_2[h]\right\}
	\frac{q(q+1)\cdots(q+m-5)}{2(m-2)!}.
\end{eqnarray}

\section{Results}
\label{results}
We start by comparing the values of packing fraction at the bifurcations from the I phase to the different
$p$-atic phases, using an extended SPT with increasing number of virial coefficients. Table \ref{table1}
presents the results. Except $B_5^*$-SPT (which is based on an extrapolated fifth virial coefficient), all of the 
theories predict a first bifurcation to the $2$-atic phase. However, the next bifurcation is invariably the
I$-8$-atic bifurcation, which shows the tendency of the HRT fluid to develop octatic correlations. Also, the
difference $\Delta\eta\equiv\eta_4-\eta_1$ between the I$-8$-atic and I$-2$-atic bifurcations tends 
to dramatically decrease when
the theory is extended with the fourth virial coefficient (cf. $\Delta\eta=0.1195$ for the standard SPT,
based on the second virial coefficient, with $\Delta\eta=0.0118$ for $B_4$-SPT, but 
with the difference
between the $B_3$-SPT and standard SPT, $\Delta\eta=0.1028$, hardly changing).
This clearly demonstrates the importance of four-particle 
correlations in this system. These correlations are involved
in the formation of very stable tetramers of triangles with square shape, which may dominate the properties of the
system at high densities.

The HRT fluid has been examined in detail by Gantapara et al. \cite{Gan} using MC simulation. This work predicts
a first-order phase transition from the I phase to a 4-atic liquid-crystal phase. Our own simulations 
\cite{MAR1} indicate that the nature of the liquid-crystal phase strongly depends on the protocol (either
compression or expansion) and starting configurations used in the simulations, to the extent that the equilibrium
configurations may exhibit strong octatic or purely tetratic correlations.
The value of packing fraction at which the I phase changes to the liquid-crystal phase was obtained to be $\eta=0.733$.
From Table \ref{table1} we can see that the $B_4$-SPT approximation is very good at predicting the correct density.
Whether the symmetry of the liquid-crystal phase is tetratic or octatic is a more delicate question that demands
further analysis. Note that the predicted values for $\eta_2\sim 1$ are large in all cases, due to the small
values of the ${\cal B}_n^{(2)}$ coefficients. This may indicate that the I-$4$-atic transition is of first order, 
as shown in Appendix \ref{first_order} in the framework of SPT.

\begin{table}[H]
	\begin{center}
\begin{tabular}{||l|c|c|c|c||} \hline\hline
Bifurcation & I-$2$-atic & I-$4$-atic & I-$6$-atic & I-$8$-atic\\
\hline
 $\eta_n$ from SPT & 0.8249 & 0.9928 & 0.9821 & 0.9444 \\
 \hline
 $\eta_n$ from $B_3$-SPT & 0.7325 & 0.9794 & 0.9328 & 0.8353\\
 \hline
 $\eta_n$ from $B_4$-SPT & 0.7281 & 0.9681 & 0.8631 & 0.7399\\
 \hline
 $\eta_n$ from $B_5^*$-SPT & 0.7255 & 0.9590 & 0.8304 & 0.7091\\
 \hline
 $\eta_n$ from $B_4$ & 1.0038 & 3.3121 & 1.8485 & 1.2049\\
 \hline
 $\eta_n$ from $B_5^*$ & 0.8715 & 2.1817 & 1.3071 & 0.9326\\
        \hline\hline
\end{tabular}
	\end{center}
        \caption{Values of the packing fractions $\eta_n$ at I-$2$-atic ($n=1$),
        I-$4$-atic ($n=2$), I-$6$-atic ($n=3$) and I-$8$-atic ($n=4$) bifurcations from the SPT,
        $B_3$-SPT and $B_4$-SPT theories, and from the truncated virial expansions
	$B_4$ and $B_5^*$.}
\label{table1}
\end{table}

To complete the picture, we have also calculated the bifurcations using truncated virial expansions instead of
the resummed virial series based on the SPT approximation. The results are shown in Table \ref{table1}. In common
with the standard Onsager-like theories for low aspect-ratio particles in two dimensions, the values of packing
fractions are absurdly large and generally unphysical for the theories based on $B_2$ and $B_3$ virial 
coefficients (not shown). 
But this is because density correlations are too grossly
represented while angular correlations are expected to be more faithfully captured by the virial coefficients.
Therefore the trends in packing fraction values as more virial coefficients are added may be relevant.
In line with the SPT-based theories, the bifurcation of the I$-$8-atic transition becomes closer to the
I$-$2-atic bifurcation as more virial coefficients are added, again demonstrating the importance of the 
higher (especially the fourth) virial coefficients to represent the structure of the
HRT fluid.

\begin{figure}[H]
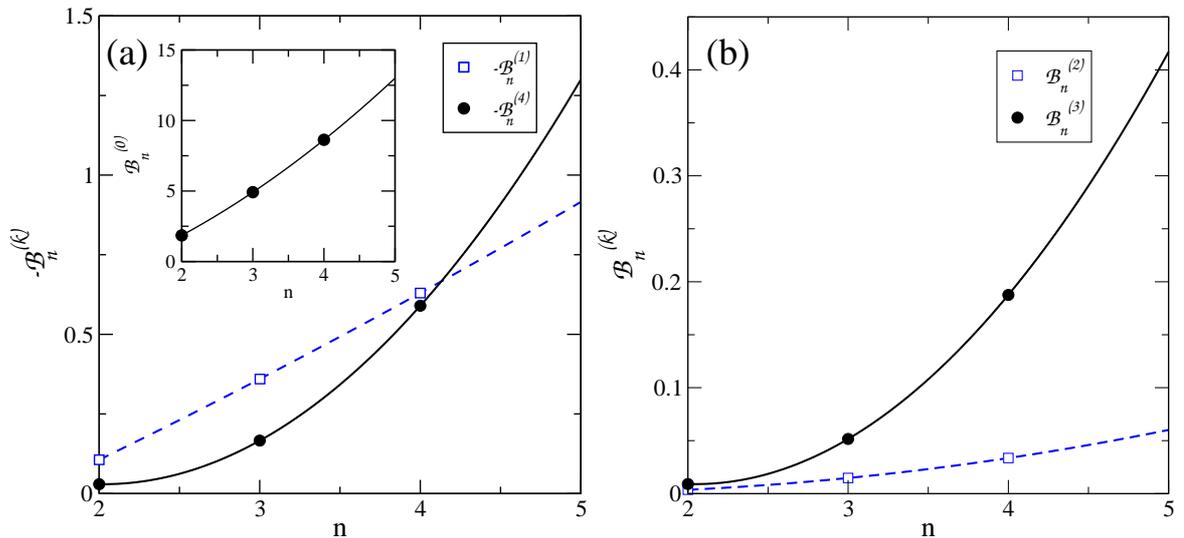

	\begin{center}
        \epsfig{file=fig1a.eps,width=3.in}
        \epsfig{file=fig1b.eps,width=3.in}
	\end{center}
	\caption{\label{extrapolated}
	Virial Fourier components ${\cal B}_n^{(k)}$ for $k=1,4$ (a) (the inset shows the case $k=0$)
        and $k=2,3$ (b) together with parabolic extrapolation to $n=5$.}
\end{figure}

To further explore the effect of higher-order virial coefficients, we have obtained extrapolated values for 
the fifth virial coefficient. Note that, even though the third- and fourth-order virial coefficients are obtained
numerically with a numerical effort which is acceptable, the fifth virial coefficient is much harder to obtain 
(especially their necessary Fourier projections 
are highly fluctuating with the relative angle between particles 
and require very detailed and costly MC integrations). Therefore we simply extrapolate
the lower-order coefficients. This is shown in Fig. \ref{extrapolated}, where the values of ${\cal B}_n^{(k)}$, for
$n=2,3$ and $4$, and $k=0,\cdots,4$, are shown. We can see that all the absolute values of the coefficients 
are increasing functions of $n$.
The figures include parabolic interpolations to the data. We take the corresponding extrapolations up to $n=5$ and
examine the consequences. Coming back to the bifurcation densities, we see from Table \ref{table1} that the trend for
$\Delta\eta$ as more virial coefficients are taken into account continues to be decreasing, actually becoming negative. This
indicates that fifth-order correlations further promote octatic ordering. This is just a trend, since the extrapolated values for
${\cal B}_4^{(k)}$, and the resulting $B_5^*$-SPT theory, need not be accurate. 
The conclusion is the same if one looks at the
truncated virial series results, Table \ref{table1}, with $\eta_2<\eta_4$ but very close. As a side comment, it is interesting
that the first two bifurcations of the truncated virial expansion of the fifth order are at physical packing fraction values (i.e.
below unity), in contrast to the lower-order expansions.

We now turn to a discussion on the thermodynamics of the different approximations by examining the equation of state in the
I phase. In the results presented above the exponent $q$ was set to a value of $2$. Now we consider flexible choices for $q$ and
examine the consequences for the equation of state in the whole range of densities. In all cases we compare with the
MC results of Gantapara et al. \cite{Gan} from the compression runs
(these data have been digitized from the original article and presented as a smooth curve to aid in visualizing the data
in comparison with the different theoretical equations of state)

\begin{figure}[H]
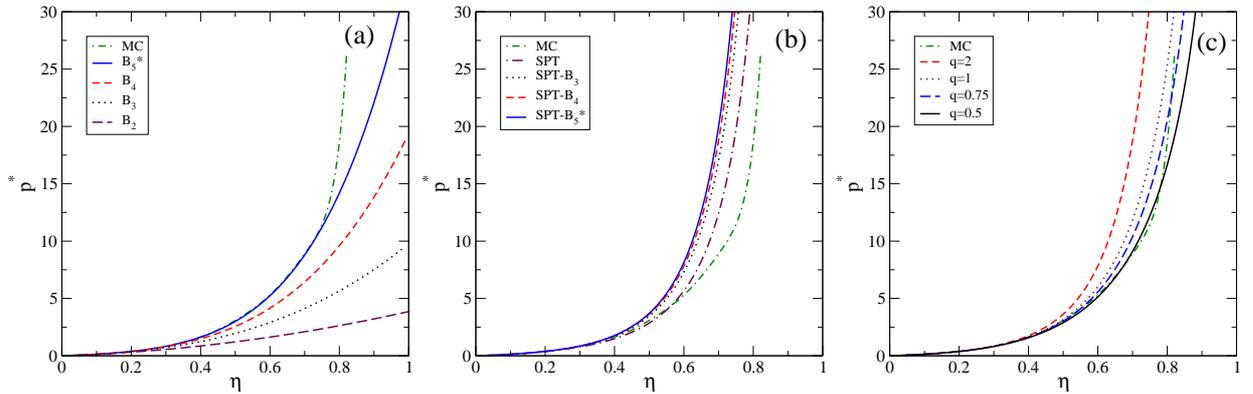

	\begin{center}
        \epsfig{file=fig2a.eps,width=2.1in}
        \epsfig{file=fig2b.eps,width=2.1in}
        \epsfig{file=fig2c.eps,width=2.1in}
	\end{center}
        \caption{(a) Equations of state from the 2nd, 3rd, 4th and 5th virial expansions. (b) Equations of state from  Eqn. (\ref{lolo})
        selecting $q=2$ and
        fixing the correct virial coefficients up to second, third, fourth and fifth orders (the later using
        its extrapolated value). (c) Equations of state from Eqn. (\ref{lolo}) using the correct $B_n$ ($n=2,\dots,4$)
        and using different values of $q$.
        }
        \label{nueva}
\end{figure}

Fig. \ref{nueva} presents the equations of state in the I phase, for the different
theories and also with different values of $q$.
Note that, according to the simulations, the I phase is stable up to 
$\eta=0.733$, but the compression results from the simulations could be taken
as reproducing a metastable I phase beyond the liquid-crystal transition. As we have shown previously \cite{MAR1},
the structure of the fluid in this regime appears to be quite complex, with different clusters of particles which give
rise to sampling problems in the simulations. Therefore, the comparison beyond this density should be taken with caution,
but we extend the density interval shown in the figures in order to assess the impact of different diverging behaviours.

We start by looking at Fig. \ref{nueva}(a), which shows the equations of state from truncated virial expansion. 
As expected the addition of more virial coefficients improve the results (note that only the ${\cal B}_n^{(0)}$ coefficients
are needed in the I phase). But it is remarkable that the expansion based on $B_5^*$ 
is very accurate in the whole density interval where the I phase is stable. 
Clearly the addition of the extrapolated fifth virial coefficient corrects almost completely the equation of state, 
which demonstrates that the extrapolation may be accurate.

Let us consider the equations of state as derived from the resummed expansions based on SPT, i.e. selecting $q=2$. 
This is shown in 
Figs. \ref{nueva}(b). Although the pressure is correctly reproduced at moderate densities by
all theories, the results are somewhat disappointing at high density when they are compared with the simulations.
This is otherwise expected, as resummed theories are known to be relatively accurate in predicting densities of phase
transitions to liquid-crystal phases, but not so in reproducing the equations of state quantitatively.

\begin{figure}[H]
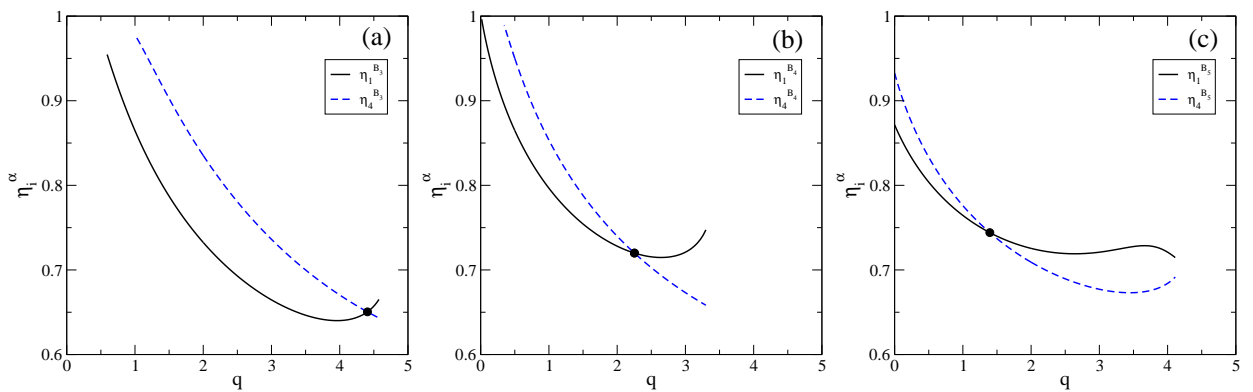

	\begin{center}
	\epsfig{file=fig3a.eps,width=2.1in}
	\epsfig{file=fig3b.eps,width=2.1in}
	\epsfig{file=fig3c.eps,width=2.1in}
	\end{center}
	\caption{Packing fractions $\eta_1(q)$ and 
	$\eta_4(q)$ resulting from (a) $B_3$-SPT, (b) $B_4$-SPT and (c) $B_5^*$-SPT theories. 
	In all cases the bifurcation condition $\chi_n(\eta)=0$ has been applied.}
	\label{last}
\end{figure}

Finally, we present an investigation which aims to explore the effect of different diverging behaviours of the equation of state
on the pressure and the bifurcation points. This is controlled by the parameter $q$ introduced in Eqn. (\ref{lolo}).
We present some results for the cases $q=0.5$, $0.75$ and $1$. Fig. \ref{nueva}(c)
show the equations of state for these cases. Clearly, as $q$ is decreased the results improve significantly. The best quantitative 
agreement is obtained for $q=0.5$.

The changes in the bifurcation points are shown in Fig. \ref{last}, which are focused on the I-$2$-atic and
and I-$8$-atic bifurcations,
represented by $\eta_1$ and $\eta_4$, respectively. In the different panels we represent the variations of both parameters
as $q$ is changed within a wide interval, for the three resummed theories incorporating the third, fourth and (extrapolated)
fifth virial coefficients. The crossover between $\eta_1$ and $\eta_4$ is induced as 
$q$ increases, but the crossover value decreases with the
order of the theory. Note that, for the $B_4$-SPT theory, the critical value of $q$ is close to $2$. Note that an optimal
equation of state requires values of $q$ which are not optimal for the bifurcation point, but the inclusion of higher
virial coefficients may improve this situation, as shown by the resummed theory based on an extrapolated fifth virial
coefficient. 

\section{Conclusions}
\label{conclusions}

In this work we have examined the effect of the high-order virial coefficients on the structure and thermodynamics of
the HRT fluid. Virial coefficients and relevant Fourier projections up to fourth order have been evaluated by MC integration.
In addition, the fifth-order virial coefficient has been extrapolated. 
When compared to simulation, the fifth-order truncated virial expansion, which requires the extrapolated value of the zeroth-order
Fourier component ${\cal B}_5^{(0)}$, gives a good approximation for
the equation of state of the I phase up to a packing fraction of $\eta\simeq 0.75$. Note that this
is slightly beyond the predicted range of stability for the I phase before the liquid-crystal phase becomes stable.
In addition, it is the lowest-order truncated virial expansion for which the I-$2$-atic and I-$8$-atic 
bifurcation points occur at 
packing fractions below unity. The resummed $B_n$-SPT approaches overestimate the pressure, but packing fractions at 
bifurcation are correct as compared with simulation for $n>2$. To investigate the impact of different diverging behaviours of the equation of state, 
we explored values of the exponent $q$ different from the standard choice $q=2$. As $q$ is decreased from $1$ to $0.5$ 
the resulting equation of state compares quite reasonably with simulation. However, the values of the bifurcation
packing fractions $\eta_1$ and $\eta_4$ deviate quite significantly from simulation, and the difference
$\Delta \eta=\eta_4-\eta_1$ increases as $q$ decreases.

As far as the resummed theories are concerned, the $B_4$-SPT theory gives $\eta_1\lesssim \eta_4$, while the $B_5^*$-SPT
theory (with an extrapolated fifth virial coefficient) already predicts $\eta_1 > \eta_4$. The crossover point is located
between $q=1$ and $1.5$, 
which in turn implies that the equation of state will be better than in the standard case.
An obvious outcome of this investigation is that minor changes in the form of the equation of state may have important 
quantitative consequences. Overall, the results presented in this article point to the necessity to include virial coefficients 
beyond the fourth if the equation of state and the I-$8$-atic bifurcation point are to be reproduced correctly. 
Unfortunately, such an effort would incur a high computational
cost, rendering the approach based on Eqn. (\ref{lolo}) impractical. Clearly, an alternative approach is needed to explain 
the behaviour of the HRT fluid and the correct symmetry of its liquid-crystal phase. 
We are inclined to believe that this theory should include the effect of particle clustering into 'superparticles', such as 
square tetramers obtained by joining four triangular monomers. These square-shaped configurations will certainly stabilize 
the tetratic phase, while other clusters such as square dimers, in equilibrium with tetramers, will produce octatic correlations.

\appendix

\section{Fourier components of virial coefficients}
\label{app}

In this section we give details on the calculation of the Fourier components of the virial coefficients $B_n[h]$,
$n=2,\dots,4$. These components are needed in the analysis of the bifurcation points from the I fluid to the $p$--atic phase. 
The virial coefficients  are in general functionals of the orientational distribution function $h(\phi)$ and can be written in 
the usual diagrammatic form as \cite{Hoover}
\begin{eqnarray}
	&&\begin{tikzpicture}
        \draw node at (-1.4,0) {$-2B_2[h]A=$};
        \filldraw[black] (0,0) circle (0.1);
        \draw[thick] (0,0)--(1,0);
        \filldraw[black] (1,0) circle (0.1);
        \draw node at (0.1,-0.2) {{\tiny 1}};
        \draw node at (0.9,-0.2) {{\tiny 2}};
        \draw node at (0.,0.3) {{\tiny ({\bf 0},0)}};
        \draw node at (1.,0.3) {{\tiny $({\bm r},\phi)$}};
\end{tikzpicture}
        \label{b2}\\
	&&\begin{tikzpicture}
        \draw node at (-1.4,0.4) {$-3B_3[h]A=$};
        \filldraw[black] (0,0) circle (0.1);
        \draw[thick] (0,0)--(1,0);
        \filldraw[black] (1,0) circle (0.1);
        \filldraw[black] (0.5,0.8) circle (0.1);
        \draw[thick] (0,0)--(0.5,0.8);
        \draw[thick] (1,0)--(0.5,0.8);
        \draw node at (0.5,0.55) {{\tiny 1}};
        \draw node at (0.75,0.15) {{\tiny 2}};
        \draw node at (0.25,0.15) {{\tiny 3}};
        \draw node at (0.5,1.05) {{\tiny ({\bf 0},0)}};
        \draw node at (1.05,-0.25) {{\tiny $({\bm r},\phi)$}};
        \draw node at (0.01,-0.25) {{\tiny $({\bm r}',\phi')$}};
\end{tikzpicture}
        \label{b3}\\
	&&\begin{tikzpicture}
        \draw node at (-1.7,0.5) {$-4B_4[h]A=\displaystyle{\frac{3}{2}}$};
       \filldraw[black] (0.,0) circle (0.1);
       \draw node at (0.15,0.85) {{\tiny 1}};
        \draw node at (0.85,0.85) {{\tiny 2}};
        \draw node at (0.85,0.15) {{\tiny 3}};
        \draw node at (0.15,0.15) {{\tiny 4}};
        \draw node at (0.,1.23) {{\tiny ({\bf 0},0)}};
        \draw node at (1.,1.23) {{\tiny $({\bm r},\phi)$}};
        \draw node at (1.,-0.23) {{\tiny $({\bm r}',\phi')$}};
        \draw node at (0.0,-0.23) {{\tiny $({\bm r}'',\phi'')$}};
        \draw[thick] (0,0)--(1,0);
        \filldraw[black] (1,0) circle (0.1);
        \draw[thick] (1,0)--(1,1);
        \filldraw[black] (1,1) circle (0.1);
        \filldraw[black] (0,1) circle (0.1);
        \draw[thick] (1,1)--(0,1);
        \draw[thick] (0,1)--(0,0);
        \draw node at (1.6,0.5) {$+\ 3$};
        \filldraw[black] (2.2,0) circle (0.1);
        \draw[thick] (2.2,0)--(3.2,0);
        \filldraw[black] (3.2,0) circle (0.1);
        \draw[thick] (3.2,0)--(3.2,1);
        \filldraw[black] (3.2,1) circle (0.1);
        \filldraw[black] (2.2,1) circle (0.1);
        \draw[thick] (3.2,0)--(3.2,1);
        \draw[thick] (3.2,1)--(2.2,1);
        \draw[thick] (2.2,1)--(2.2,0);
        \draw[thick] (2.2,1)--(3.2,0);
        \draw node at (3.8,0.5) {$+\ \displaystyle{\frac{1}{2}}$};
       \filldraw[black] (4.5,0) circle (0.1);
       \draw[thick] (4.5,0)--(5.5,0);
        \filldraw[black] (5.5,0) circle (0.1);
        \draw[thick] (4.5,0)--(4.5,1);
        \filldraw[black] (4.5,1) circle (0.1);
        \filldraw[black] (5.5,1) circle (0.1);
        \draw[thick] (4.5,0)--(5.5,1);
        \draw[thick] (4.5,1)--(5.5,1);
        \draw[thick] (5.5,0)--(5.5,1);
        \draw[thick] (4.5,1)--(5.5,0);
\end{tikzpicture}
        \label{b4}
\end{eqnarray}

Nodes and bonds label particles and Mayer functions, respectively. The reference system is located at node 1 which means that its 
position and angle are $({\bm r}_1,\phi_1)=({\bf 0},0)$. Therefore the nodes labelled as 2, 3 and 4 (in a clockwise direction) have 
relative coordinates ${\bm r}_2-{\bm r}_1\equiv {\bm r}$, ${\bm r}_3-{\bm r}_1\equiv{\bm r}'$, ${\bm r}_4-{\bm r}_1\equiv{\bm r}''$, $\phi_2-\phi_1\equiv \phi$, 
$\phi_3-\phi_1\equiv \phi'$, $\phi_4-\phi_1\equiv\phi''$. The spatial and angular integrations with respect to ${\bm r}_1$ and 
$\phi_1$ in $B_n[h]$ can then be performed trivially. In particular, a factor equal to the total area $A$ cancels out,
see Eqs. (\ref{b2})-(\ref{b4}). Integration over $\phi_1$ allows to define the following angular functions:
\begin{eqnarray}
        &&\Psi_2(\phi)\equiv 
	\int_0^{2\pi}d\phi_1 h(\phi_1)h(\phi_1+\phi), \label{psi2}\\
        &&\Psi_3(\phi,\phi')\equiv 
	\int_0^{2\pi}d\phi_1 h(\phi_1)h(\phi_1+\phi)h(\phi_1+\phi'), 
	\label{psi3}\\
        &&\Psi_4(\phi,\phi',\phi'')\equiv 
        \int_0^{2\pi}d\phi_1 h(\phi_1)h(\phi_1+\phi)h(\phi_1+\phi')
	h(\phi_1+\phi''), \label{psi4}
\end{eqnarray}
Close to the I-$2n$-atic bifurcation point we can approximate $h(\phi)$, up to first order, as 
\begin{eqnarray}
	h(\phi)\simeq \frac{1}{2\pi}\left[1+h_n \cos(2n\phi)\right],\quad n=1,\dots,4.
\end{eqnarray}
To lowest order of the angular functions become
\begin{eqnarray}
        &&\Psi_2(\phi)=
         \frac{1}{2\pi}\left\{
                 1+\frac{h_{n}^2}{2}
			 \cos(2n\phi)\right\}, \label{aprox2}\\
        &&\Psi_3(\phi,\phi')=
         \frac{1}{(2\pi)^2}\left\{
                 1+\frac{h_{n}^2}{2}\left[
                         \cos(2n\phi)+\cos(2n\phi')+\cos(2n\phi)\cos(2n\phi')\right]\right\},
	 \label{aprox 3}\\
        &&\Psi_4(\phi,\phi',\phi'')=
         \frac{1}{(2\pi)^3}\left\{
                 1+\frac{h_{n}^2}{2}\left[
                         \cos(2n\phi)+\cos(2n\phi')+\cos(2n\phi'')
                         \right.\right.\nonumber\\
                         &&\left.\left. +\cos(2n\phi)\cos(2n\phi')+\cos(2n\phi)\cos(2n\phi'')
                         +\cos(2n\phi')\cos(2n\phi'')\right]\right\}.
	 \label{aprox4}
\end{eqnarray}
The spatial integrals involved in $B_n[h]$ can be defined to be angular kernels, 
\begin{eqnarray}
        &&{\cal K}_2(\phi)=-\frac{1}{2}\int d{\bm r}f({\bm r},\phi), 
	\label{kk2}\\
	&&{\cal K}_3(\phi,\phi')=-\frac{1}{3}\int d{\bm r}\int d{\bm r}' f({\bm r},\phi)f({\bm r}',\phi') f({\bm r}-{\bm r}',\phi-\phi'),\label{kk3}\\
        &&{\cal K}^{(1)}_4(\phi,\phi',\phi'')=-\frac{3}{8}\int d{\bm r}\int d{\bm r}'\int d{\bm r}'' f({\bm r},\phi)f({\bm r}'',\phi'') f({\bm r}-{\bm r}',\phi-\phi')\nonumber\\
	&&\hspace{7cm}\times f({\bm r}'-{\bm r}'',\phi'-\phi''). \label{kk4a}\\
        &&{\cal K}^{(2)}_4(\phi,\phi',\phi'')=-\frac{3}{4}\int d{\bm r}\int d{\bm r}'\int d{\bm r}'' f({\bm r},\phi)f({\bm r}',\phi') f({\bm r}'',\phi'') f({\bm r}-{\bm r}',\phi-\phi')\nonumber\\
	&&\hspace{7cm}\times f({\bm r}'-{\bm r}'',\phi'-\phi'').\label{kk4b}\\
        &&{\cal K}^{(3)}_4(\phi,\phi',\phi'')=-\frac{1}{8}\int d{\bm r}\int d{\bm r}'\int d{\bm r}'' f({\bm r},\phi)f({\bm r}',\phi') f({\bm r}'',\phi'') f({\bm r}-{\bm r}',\phi-\phi')\nonumber\\
	&&\hspace{7cm}\times f({\bm r}'-{\bm r}'',\phi'-\phi'') f({\bm r}-{\bm r}'',\phi-\phi''),
	\label{kk4c}
\end{eqnarray}
Note that ${\cal K}_2(\phi)$ is just half the excluded area. The superindex $m$ in the definition of 
${\cal K}_4^{(m)}(\cdots)$ labels the empty square-diagram ($m=1$),
the square-diagram with one diagonal ($m=2$), and the square-diagram with two diagonals ($m=3$), respectively.
To implement the bifurcation analysis we need the following Fourier components of these kernels:
\begin{eqnarray}
        &&{\cal K}_{2,n}=\int_0^{2\pi}d\phi\cos(2n\phi){\cal K}_2(\phi), 
	\label{k2}\\
        &&{\cal K}_{3,n,m}=\int_0^{2\pi}d\phi\cos(2n\phi)\int_0^{2\pi} d\phi'\cos(2m\phi')
	{\cal K}_3(\phi,\phi'), \label{k3}\\
        &&{\cal K}^{(k)}_{4,n,m,l}=\int_0^{2\pi}d\phi\cos(2n\phi)\int_0^{2\pi} d\phi'\cos(2m\phi')\int_0^{2\pi}d\phi''
	\cos(2l\phi''){\cal K}^{(k)}_4(\phi,\phi',\phi''), \label{k4}
\end{eqnarray}
Finally, the virial coefficients $B_n[h]$ can be approximated, close to the bifurcation point, by using Eqns. (\ref{b2})-(\ref{b4}), 
(\ref{aprox2})-(\ref{aprox4}), and (\ref{kk2})-(\ref{kk4c}), as 
\begin{eqnarray}
	\hspace{-1.2cm}B_2[h]&=&\prod_{k=1}^2\left(\int_0^{2\pi} d\phi_k h(\phi_k)\right)
        {\cal K}_2(\phi)
        =\int_0^{2\pi} d\phi
        \Psi_2(\phi){\cal K}_2(\phi)=B_2^{(0)}+\frac{h_n^2}{2}B_2^{(n)},\nonumber\\
	\hspace{-1.2cm}B_3[h]&=&\prod_{k=1}^3\left(\int_0^{2\pi} d\phi_k h(\phi_k)\right)
        {\cal K}_3(\phi,\phi')\nonumber\\
	&=&\int_0^{2\pi} d\phi\int_0^{2\pi}d\phi'
        \Psi_3(\phi,\phi'){\cal K}_3(\phi,\phi') =B_3^{(0)}+\frac{h_n^2}{2}B_3^{(n)}\nonumber\\
	\hspace{-1.2cm}B_4[h]&=&\prod_{k=1}^4\left(\int_0^{2\pi} d\phi_k h(\phi_k)\right)
        \left(\sum_{m=1}^3{\cal K}^{(m)}_4(\phi,\phi',\phi'')\right)\nonumber\\
	&=&\int_0^{2\pi} d\phi\int_0^{2\pi}d\phi'
        \int_0^{2\pi} d\phi'' \Psi_4(\phi,\phi',\phi'') \left(\sum_{m=1}^3{\cal K}^{(m)}_4(\phi,\phi',\phi'')\right)
	=B_4^{(0)}+\frac{h_n^2}{2}B_4^{(n)},
\end{eqnarray}
with
\begin{eqnarray}
	&&B_2^{(0)}=\frac{{\cal K}_{2,0}}{2\pi},\hspace{0.4cm}
	B_2^{(n)}=\frac{{\cal K}_{2,n}}{2\pi}\nonumber\\
	&&B_3^{(0)}=\frac{{\cal K}_{3,0,0}}{(2\pi)^2},\hspace{0.4cm}B_3^{(n)}=\frac{1}{(2\pi)^2}\left(2{\cal K}_{3,n,0}+{\cal K}_{3,n,n}\right)\nonumber\\
	&&B_4^{(0)}=\frac{1}{(2\pi)^3}\sum_{m=1}^3 {\cal K}^{(m)}_{4,0,0,0},\nonumber\\
	&&B_4^{(n)}=\frac{1}{(2\pi)^3}\left\{\sum_{m=1}^2 \left[2 {\cal K}^{(m)}_{4,n,0,0}+{\cal K}^{(m)}_{4,0,n0}+
        2{\cal K}^{(m)}_{4,n,n,0}+{\cal K}^{(m)}_{4,n,0,n}\right]
        +3\left[{\cal K}^{(3)}_{4,n,0,0}+{\cal K}^{(3)}_{4,n,n,0}\right]\right\}.\nonumber\\
	&&
\end{eqnarray}
From these approximations we define the scaled Fourier components of the virial coefficients:
	\begin{eqnarray}
        {\cal B}_k^{(0)}\equiv \frac{B_k^{(0)}}{a^{k-1}}-1,\quad
		{\cal B}_{k}^{(n)}\equiv \frac{B_k^{(n)}}{a^{k-1}}, \label{scaled}
	\end{eqnarray}

The second-order coefficients ${\cal B}_2^{(n)}$ can be calculated analytically:
\begin{eqnarray}
        {\cal B}_2^{(n)}=-\frac{8}{\pi(4n^2-1)} 
        \cos^2\left[\frac{(2n-1)\pi}{8}\right]
        \cos^2\left[\frac{(2n+1)\pi}{8}\right],
	\label{analytic}
\end{eqnarray}
The remaining coefficients have to be computed numerically. We have used MC integration for the spatial integrals and 
Gaussian quadratures for the angular integrals, using special tricks to deal with the rapidly varying trigonometric
functions of high index. $10^{8}$ configurations were used to evaluate the spatial
integrals. The results are collected in Table \ref{rasa}. 

\begin{table}[H]
	\begin{center}
\begin{tabular}{||c|c|c|c|c|c||} \hline\hline
	$k$ & ${\cal B}_k^{(0)}$ & ${\cal B}_k^{(1)}$ & ${\cal B}_k^{(2)}$ & ${\cal B}_k^{(3)}$ & ${\cal B}_k^{(4)}$\\
\hline
	$2$  &$1.8552$&$-0.1061$&$-0.0036$&$-0.0091$&$-0.0294$ \\
 \hline
	$3$  &$4.9158$&$-0.3597$&$-0.0148$&$-0.0516$&$-0.1662$ \\
 \hline
	$4$  &$8.6307$&$-0.6296$&$-0.0336$&$-0.1876$&$-0.5897$\\
        \hline\hline
\end{tabular}
	\end{center}
	\caption{Values of the coefficients ${\cal B}_k^{(n)}$, obtained analytically for $k=2$ from Enq. (\ref{analytic}),
	and numerically from MC integration and Gaussian quadrature for $k>2$.} 
\label{rasa}
\end{table}

\section{Relative stability of the 4--atic phase}
\label{first_order}

\begin{figure}
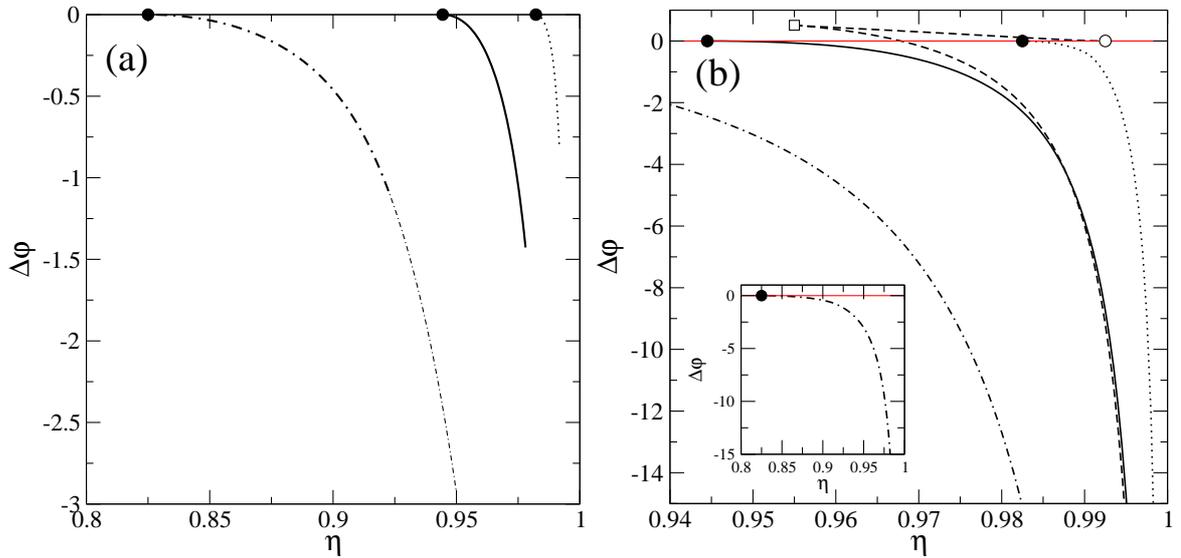

                        \epsfig{file=fig4a.eps,width=3.in}
                        \epsfig{file=fig4b.eps,width=3.in}
                        \caption{Free-energy differences $\Delta\varphi\equiv \varphi_{\alpha}
                        -\varphi_{I}$, with $\alpha$= 2--atic (dot-dashed), 8--atic (solid), 4--atic (dashed) and
			6--atic (dotted), as a function of packing fraction $\eta$, obtained from (a) 
			the Fourier-coefficient method, and (b) the one-parameter minimizations.
                        In (a) the dot-dashed line with a shorter step is a simple extrapolation
                        of the $2$--atic branch to higher densities. The inset
                        in (b) shows the complete $2$--atic branch. Filled circles in (a) and (b) show
                        the bifurcation points at second order transitions, while the open circle
                        in (b) indicates the first-order counterpart in the 4--atic branch. The
                        open square is the location of the first metastable solution with
                        4--atic symmetry.}
                        \label{fig_nueva}
                \end{figure}

For those $n$-th virial theories which predict an I-2--atic bifurcation below the I-8--atic 
one may wonder which of the following scenarios takes place at densities above
the I-8--atic bifurcation: (i) The 8--atic and 2--atic free-energy branches cross each other at some density, or (ii)
the 2--atic branch continues to be the lowest one. 
In order to investigate this point, we have minimised the
B$_2$-SPT functional, considering a subset of Fourier coefficients $\{h_k\}$ with $k=4j$:
\begin{eqnarray}
        h(\phi)=\frac{1}{2\pi}\left[1+\sum_{k=0}^{n_{\rm max}} h_k \cos(2k\phi)\right].
\end{eqnarray}
This choice gives a distribution $h(\phi)$ with perfect 8--atic symmetry. A free-energy branch was generated
for a density interval starting at the I-8--atic bifurcation point and up
to densities such that the 8--atic order parameter is $Q_8\simeq 0.97$ (we checked that with this condition
the truncated Fourier series still gives correct results).
The 2--atic branch was also calculated up to densities such that $Q_2\simeq 0.97$.
Fig. \ref{fig_nueva}(a) shows the free-energy differences $\Delta\varphi\equiv \varphi_{\rm \alpha}-\varphi_{\rm I}$
($\alpha=$2,6,8--atic) between $\alpha$ and I phases calculated from their respective bifurcation points.
As expected, the 8--atic branch is always metastable. 
However, the results indicate that the first scenario above can be discarded, as the difference between the 8--atic and 2--atic 
free-energy branches is huge (note that the latter bifurcates from the isotropic at a much lower density). 
The situation is even worse in the case of the metastable 6--atic phase, as it bifurcates at even higher packing fractions
(see Table \ref{table1}). We note that the 4--atic phase also bifurcates from the I phase but,
unlike the 2--atic, 8--atic and 6--atic phases, it does so via a first-order transition. This can be demonstrated, 
not via Fourier-amplitude minimization (we were unable to obtain a metastable 4--atic solution with the proper
restrictions over $\{h_k\}$), but from a simple one-parameter minimization of the orientation distribution,
                \begin{equation}
                        h(\phi)=\dfrac{e^{\lambda \cos(2n\phi)}}{2\pi I_0(\lambda)},\hspace{0.6cm}
			\phi\in[0,2\pi],\hspace{0.6cm} n=1,\cdots,4.
                \end{equation}
		$I_0(x)$ is the zeroth-order modified Bessel function of the first kind, and
		$\lambda$ is a variational parameter.
                The results are shown in Fig. \ref{fig_nueva}(b):
                from the bifurcation point (open circle in the figure),
                located at a packing fraction $\eta=0.9928$
                (see Table \ref{table1}), an unstable 4--atic branch departs towards lower densities,
                with a free energy higher than that of the I phase. This branch
                terminates at $\eta\approx 0.955$ (open square in the figure), where the $4$-atic phase
                becomes metastable for the first time, with a high order parameter $Q_4$.
                From this point a second T branch develops towards higher densities,
                $\Delta\varphi$ eventually becoming negative at $\eta\simeq 0.968$,
                indicating that the $4$-atic phase is more stable than the I phase.
                This is the usual scenario for a first-order phase transition.
                When $\eta$ further increases from this value, the $4$-atic free-energy branch
                also crosses the $8$-atic branch at $\eta\simeq 0.988$.
                In any case, the free energy of the 2--atic phase has by far the lowest value,
                as can be seen in Fig. \ref{fig_nueva}. The crossing of the $4$-atic and $8$-atic branches
                at high packing fractions is interesting. It can be understood by invoking
                the limit value of the scaled second-virial coefficient as
                $\lambda\to\infty$:
                \begin{eqnarray}
			&&\tilde{b}_2^{(n)}\equiv \lim_{\lambda\to\infty} b_2[h]\nonumber\\
                        &&=\frac{1}{2an}\left[\frac{A_{\rm excl}(0)+A_{\rm excl}(\pi)}{2}
                        +\sum_{k=1}^{n-1}A_{\rm excl}\left(\frac{k\pi}{n}\right)\right]-1.
                        \nonumber\\
                \end{eqnarray}
                Inserting the known analytic expression for the excluded area $A_{\rm excl}(\phi)$, we obtain
                $\tilde{b}_2^{(1)}<\tilde{b}_2^{(2)}\lesssim \tilde{b}_2^{(4)}<\tilde{b}_2^{(3)}$ for the
                2--atic ($n=1$), 4--atic ($n=2$), 6--atic ($n=3$) and 8--atic ($n=4$) symmetries. 
		This explains
                the reason for the crossing behavior: the double-averaged excluded area with
                respect to $h(\phi)$ for the 4--atic symmetry, although similar in magnitude,
                is lower than that obtained for the 8--atic symmetry. Obviously this occurs only
                at very high densities, when the orientational order is almost perfect
                and the above asymptotic expression can be justified.
		The fact that the free energy of the $8$-atic phase is lower than that 
		of the $4$-atic phase
                at lower densities implies that the opposite behavior is true when the
                orientational distribution function is less sharply peaked.
                $n$th-order virial theories with an I-8--atic bifurcation below the I-2--atic one,
		with the free-energy of the former below that of the latter, are expected to
		support the second scenario: After the I-8--atic bifurcation at high packing fractions,
		the 4--atic energy branch crosses the 8--atic one, and a $8$-atic--$4$-atic first-order
		phase transition takes place. However, if the crossing point
                is relatively close to the I-8--atic bifurcation, the two-phase coexistence might 
		involve the I and the $4$-atic phases, as shown by the MC simulations.

\acknowledgements

Financial support from Grant No. PID2021-126307NB-C21 (MCIU/AEI/FEDER,UE) is acknowledged.

\end{document}